\def\year{2018}\relax
\documentclass[letterpaper]{article} 
\usepackage{aaai18}  
\usepackage{times}  
\usepackage{helvet}  
\usepackage{courier}  
\usepackage{url}  
\usepackage{graphicx}  
\usepackage{booktabs}
\usepackage{float}
\restylefloat{table}
\usepackage[T2A,T1]{fontenc}
\usepackage[utf8]{inputenc}
\usepackage[russian,english]{babel}
\usepackage [autostyle, english = american]{csquotes}
\MakeOuterQuote{"}
\usepackage{balance}

\begin{document}
%
\author{Jane Im,\textsuperscript{$\dag$}
Eshwar Chandrasekharan,\textsuperscript{$\dag\dag$}
Jackson Sargent,\textsuperscript{$\dag$}
Paige Lighthammer,\textsuperscript{$\dag$}\\
\Large{\textbf{Taylor Denby}},\textbf{\textsuperscript{$\dag$}}
\Large{\textbf{Ankit Bhargava}},\textbf{\textsuperscript{$\dag$}}
\Large{\textbf{Libby Hemphill}},\textbf{\textsuperscript{$\dag$}}
\Large{\textbf{David Jurgens}},\textbf{\textsuperscript{$\dag$}}
\Large{\textbf{Eric Gilbert}}\textbf{\textsuperscript{$\dag$}}
\\
\textsuperscript{$\dag$}{University of Michigan}\\
\textsuperscript{$\dag\dag$}{Georgia Institute of Technology}\\
imjane@umich.edu,
eshwar3@gatech.edu, 
jacsarge@umich.edu,
paigeal@umich.edu,\\
tdenby@umich.edu,
abharga@umich.edu,
libbyh@umich.edu,
jurgens@umich.edu, eegg@umich.edu
}
\title{Still Out There: Modeling and Identifying Russian Troll Accounts on Twitter}
\maketitle

\begin{abstract}
There is evidence that Russia's Internet Research Agency attempted to interfere with the 2016 U.S. election by running fake accounts on Twitter---often referred to as “Russian trolls”. In this work, we: 1) develop machine learning models that predict whether a Twitter account is a Russian troll within a set of 170K control accounts; and, 2) demonstrate that it is possible to use this model to find active accounts on Twitter still likely acting on behalf of the Russian state. 
Using both behavioral and linguistic features, we show that it is possible to distinguish between a troll and a non-troll with a precision of 78.5\% and an AUC of 98.9\%, under cross-validation. Applying the model to out-of-sample accounts still active today, we find that up to 2.6\% of top journalists' 
mentions are occupied by Russian trolls. These findings imply that the Russian trolls are very likely still active today. Additional analysis shows that they are not merely software-controlled bots, and manage their online identities in various complex ways. Finally, we argue that if it is possible to discover these accounts using externally-accessible data, then the platforms---with access to a variety of private internal signals---should succeed at similar or better rates.

\end{abstract}

\section{Introduction}
\begin{figure*}[t]
    \centering
    \includegraphics[width=2\columnwidth]{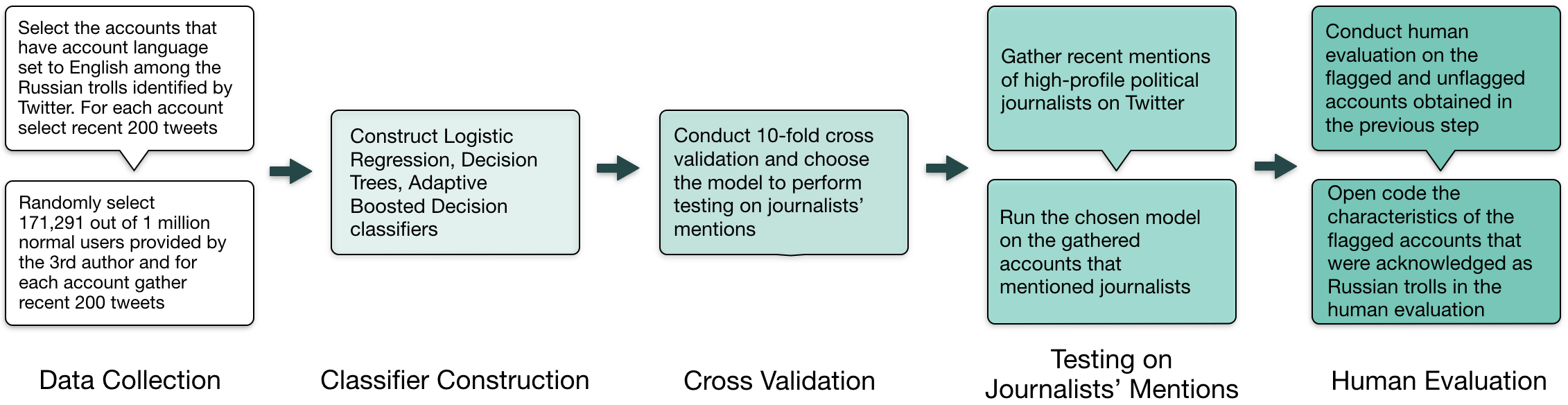}
    \caption{Flowchart illustrating the steps of our research pipeline.}
    \label{fig:steps}
\end{figure*}

It is widely believed that Russia's Internet Research Agency (IRA) tried to interfere with the 2016 U.S. election as well as other elections by running fake accounts on Twitter---often called the "Russian troll" accounts \cite{gorodnichenko2018social,ferrara2017disinformation,stella2018bots}.
This interference could have immense consequences 
considering the viral nature of some tweets \cite{mustafaraj2010obscurity,metaxas2012social}, the number of users exposed to Russian trolls' content \cite{isaac2017russian,spangher2018analysis}, and the critical role social media have played in past political campaigns \cite{cogburn2011networked}. In this paper, we develop models on a  dataset of Russian trolls active on Twitter during the 2016 U.S. elections to predict currently active Russian trolls. We construct machine learning classifiers using profile elements, behavioral features, language distribution, function word usage, and linguistic features, on a highly unbalanced dataset of Russian troll accounts (2.2K accounts, or 1.4\% of our sample) released by Twitter\footnote{\url{https://about.twitter.com/en_us/values/elections-integrity.html#data}} and ``normal'', control accounts (170K accounts, or 98.6\% of our sample) collected by the authors. (See Figure \ref{fig:steps} for a visual overview of the process used in this work.) Our goals are to determine whether ``new'' trolls can be identified by models built on ``old'' trolls and to demonstrate that troll detection is both possible and efficient, even with ``old'' data. 

We find that it is possible to disambiguate between a Russian troll account and a large number of these randomly selected control accounts among users. 
One model, a simple logistic regression, achieves a precision of 78.5\% and an AUC of 98.9\%. Next we asked whether it was possible to use the model trained on past data to unmask Russian trolls currently active on Twitter (see Figure \ref{fig:flagged_account_reply} for an example)? The logistic regression is attractive in this context as its simplicity seems most likely to generalize to out-of-sample data. Toward that end, we apply our classifier to Twitter accounts that mentioned high-profile journalists in late 2018. We find the computational model flags 3.7\% of them as statistically likely Russian trolls and find reasonable agreement between our classifier and human labelers.

Our model allows us to estimate the activity of trolls.  As a case study, we estimate the activity of suspected Russian troll accounts engaging in one type of adversarial campaign: engaging with prominent journalists.  
Since we have no way of truly knowing which of these model-identified accounts are truly Russian trolls---perhaps only the IRA  knows this---we perform a secondary human evaluation in order to establish consensus on whether the model is identifying validly suspicious accounts.
%
Our human evaluation process suggests that roughly 70\% 
of these model-flagged accounts---all of them still currently active on Twitter---are highly likely to be Russian trolls.  As a result, we estimate that Russian trolls occupy 2.6\% of the mentions of high-profile journalists' mentions.  Moreover, we find that in contrast with some prevailing narratives surrounding the Russian troll program, the model-flagged accounts do not score highly on the well-known Botometer scale \cite{davis2016botornot}, indicating that they are not simply automated software agents.

Finally, we perform an exploratory open coding of the identity deception strategies used by the currently active accounts discovered by our model. For instance, some pretend to be an American mother or a middle-aged white man via profile pictures and descriptions, but their tweet rates are abnormally high, and their tweets revolve solely around political topics. 


This paper makes the following three contributions, building on an emerging line of scholarship around the Russian troll accounts \cite{stewart2018examining,spangher2018analysis,griffin2018unsupervised,zannettou2018let,boatwrighttroll,boyd2018characterizing}.  First, we show that it is possible to separate Russian trolls from other accounts in the data previous to 2019, and that this computational model is still accurate on 2019 data. As a corollary, we believe this work establishes that a large number of Russian troll accounts are likely to be currently active on Twitter. Second, we provide our model to the research community.\footnote{URL available after blind review.} This will enable other researchers to study their own questions about the trolls, such as "What are their objectives?" and "How are they changing over time?" Third, we find that accounts flagged by our model as Russian trolls are not merely bots but use diverse ways to build and manage their online identities. Finally, we argue that if it is possible to discover these accounts using externally-accessible data, then the social platforms---with access to a variety of private, internal signals---should succeed at similar or better rates at finding and deactivating Russian troll accounts.


\section{Related Work}

First, we review what is known about Russian's interference in Western democracies via online campaigns, and then move on to the emerging work on these 2016 election related Russian trolls themselves. We conclude by discussing work on social bots, and by reviewing theories of online deception that inform the quantitative approaches in this paper.

\subsection{Russia's Interference on Political Campaigns}
While state-level online interference in democratic processes is an emerging phenomenon, new research documents Russia's online political manipulation campaigns in countries other than the United States. For instance, previous work has shown that a high volume of Russian tweets were generated a few days before the voting day in the case of the 2016 E.U. Referendum (Brexit Referendum), and then dropped afterwards \cite{gorodnichenko2018social}. Furthermore, it is suspected that Russia is behind the MacronLeaks campaign that occurred during the 2017 French presidential elections period \cite{ferrara2017disinformation}, as well as the Catalonian referendum \cite{stella2018bots}.

\subsection{Emerging Work on the 2016 Russian Trolls}
While a brand new area of scholarship, emerging work has examined the datasets of Russian trolls released by Twitter.
Researchers from Clemson University identified five categories of trolls and argued the behavior between these categories were radically different \cite{boatwrighttroll}. This was especially marked for left- and right-leaning accounts (the dataset contains both). For instance, the IRA promoted more left-leaning content than right-leaning on Facebook, while right-leaning Twitter handles received more engagement. \cite{spangher2018analysis}.

New work has looked at how the Russian troll accounts were retweeted in the context of the \#BlackLivesMatter movement \cite{stewart2018examining}---a movement targeted by the trolls. The retweets were divided among different political perspectives and the  trolls took advantage of this division. There is some disagreement about how predictable the Russian trolls are. Griffin and Bickel (2018) argue that the Russian trolls are composed of accounts with common but customized behavioral characteristics that can be used for future identification
\cite{griffin2018unsupervised}, while other work has shown that the trolls' tactics and targets change over time, implying that the task of automatic detection is not simple \cite{zannettou2018let}. Finally, the Russian trolls show unique linguistic behavior as compared to a baseline cohort \cite{boyd2018characterizing}.

\subsubsection{Users Who Interact with the Trolls.} Recent work has also examined the users who interact with the Russian trolls on Twitter. For example, misinformation produced by the Russian trolls was shared more often by conservatives than liberals on Twitter \cite{badawy2018analyzing}. Models can predict which users will spread the trolls' content by making use of political ideology, bot likelihood scores, and activity-related account metadata \cite{badawy2018falls}. 

\subsubsection{Measuring the Propaganda's Effect.} Researchers have also worked to understand the influence of the Russian trolls' propaganda efforts on social platforms by using Facebook's ads data, IRA related tweets on Twitter, and log data from browsers. 1 in 40,000 internet users were exposed to the IRA ads on any given day, but there was variation among left and right-leaning content \cite{spangher2018analysis}.
Furthermore, the influence of the trolls have been measured in platforms like Reddit, Twitter, Gab, and 4chan’s Politically Incorrect board (/pol/) using Hawkes Processes \cite{zannettou2018let}.

\subsection{Bots}
While some bots are built for helpful things such as auto-replies, bots can also often be harmful, such as when they steal personal information on social platforms \cite{ferrara2016rise} and spread misinformation \cite{shao2017spread,gorodnichenko2018social}. Previous research has shown that bots largely intervened with the 2016 election. For instance, bots were responsible for millions of tweets right before the 2016 election \cite{bessi2016social}. This was not the first time, as a disinformation campaign was coordinated through bots before the 2017 French presidential election \cite{ferrara2017disinformation}. Current attempts to detect bots include systems based on social network information, systems based on crowdsourcing and human intelligence, and machine-learning methods using indicative features \cite{ferrara2016rise}. However, previous findings show it is becoming harder to filter out bots due to their sophisticated behavior \cite{subrahmanian2016darpa}, such as posting material collected from the Web at predetermined times. 

\subsection{Deception and Identity Online}
Russian trolls tried to mask their identities on Twitter, for instance pretending to an African-American activists supporting \#BlackLivesMatter \cite{arif2018acting}. Seminal research has shown the importance of identities vary by online communities \cite{donath2002identity}. For example, the costliness of faking certain social signals is related to their trustworthiness \cite{donath2002identity}, an insight that we use to compose quantitative features. The importance and salience of identity signals (and possible deception through them) extend to nearly all social platforms. Online dating site users attend to small details in others' profiles and are careful when crafting their own profiles, since fewer cues meant the importance of the remaining ones were amplified \cite{ellison2006managing}. MySpace users listed books, movies, and TV shows in profiles to build elaborate taste performances in order to convey prestige, differentiation, or aesthetic preference \cite{liu2007social}. And on Twitter, users manage their self-presentation both via profiles and ongoing conversations \cite{marwick2011tweet}.

\begin{figure}[t!]
    \centering
    \includegraphics[width=0.8\columnwidth]{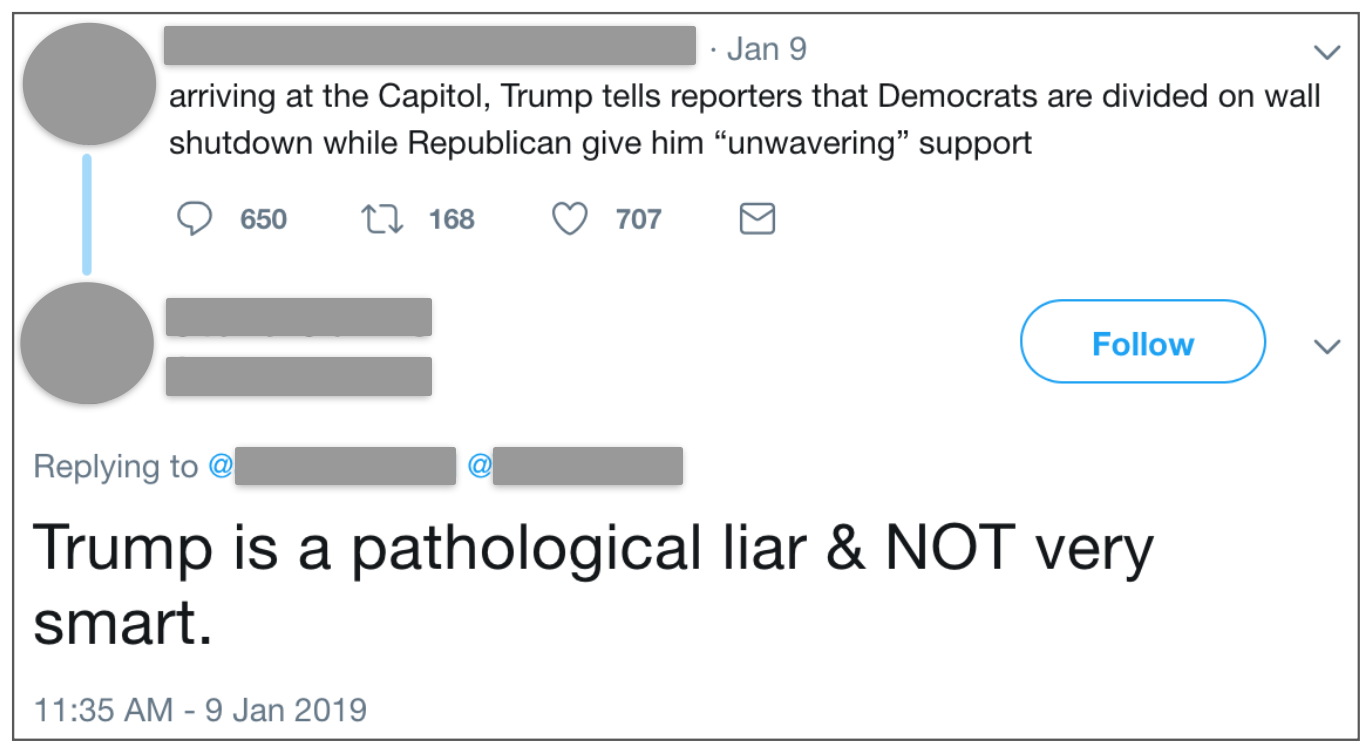}
    \caption{Example of a flagged account replying back to a high-profile journalist on Twitter.}
    \label{fig:flagged_account_reply}
\end{figure}

\section{Data} 
\label{data_section}

To model and identify potential Russian Trolls on Twitter, we first construct a large dataset of both known Russian troll accounts and a control cohort of regular users.

\subsection{Russian Trolls}
\label{russian_troll_tweets}

The suspected Russian interference in the 2016 US presidential election led to multiple federal and industry investigations to identify bad actors and analyze their behavior \cite{jensen2018russian}. %
As a part of these efforts, Twitter officially released a new dataset of 3,841 accounts believed to be connected to the Internet Research Agency (IRA).  This dataset contains features such as profile description, account creation date, and poll choices. In our paper, we used the Russian troll accounts from this new dataset for our analysis, and  model construction.

Out of the 3,841 accounts, we focus on the 2,286 accounts whose users selected English as their language of choice.  This choice was due to our goal of distinguishing a Russian troll trying to imitate a normal US user from a US user, of which the vast majority speak only English.  However, we note that despite a user selecting English, users may still tweet occasionally in other languages.
We use the most recent 200 tweets from each troll account to form a linguistic sample. This allows us to directly compare the trolls with other Twitter users, whose tweets are collected via a single Twitter API call, which only provides 200 tweets.
In total 346,711 tweets from the Russian troll dataset were used to construct the classifiers.

\subsection{Control Accounts}

To contrast with troll behavior, we construct a control dataset of users whose behavior is expected to be typical of US accounts.  The initial control accounts are drawn from a historical 10\% sample of Twitter data and then iteratively refined as by geography and activity time. To ensure geographic proximity in the US, the total variation method \cite{compton2014geotagging} is used to geolocate all users. 

We then took a random sample of US-located users and ensured they tweeted at least 5 times between 2012-2017, to match the tweet activity times in the Russian troll cohort.  
We then randomly sampled 171,291 of these accounts, which we refer to as \textit{control accounts} in the rest of the paper.
%
This creates a substantial class imbalance, with 98.6\% control accounts to 1.4\% Russian troll accounts; this imbalance matches real-world expectations that such troll accounts are relatively rare (though it is difficult to know a priori exactly how rare). 
For each control account, the most recent 200 tweets are collected (due to API restrictions), for a total dataset size of 29,960,070 tweets.  The total dataset is summarized in Table~\ref{table:dataset}. 
%

\begin{table}[t!]
\small
\begin{tabular}{@{}l l l@{}}
\textbf{}              & \textbf{Russian Trolls} & \textbf{Control Accounts} \\\midrule
Total \# of Accounts &  2,286 (1.4\%) & 171,291 (98.6\%) \\ 
Total \# of Tweets & 346,711 & 29,960,070 \\ \midrule
Avg Account Age (days)& 1679.2& 2734.5\\ 
Avg \# of Followers & 1731.8 & 1197\\ 
Avg \# of Following &  952 & 536.5\\     
\end{tabular}
\caption{Description of the dataset used to construct models.}
\label{table:dataset}
\end{table}

\subsection{Journalists' Mentions}\label{journalist_mentions}

Recent research at the intersection of social media and journalism confirms that journalists use Twitter as a source of tips, especially for breaking news \cite{Willnat2018-ll,Swasy2016-dn}. If the trolls' goals are to influence the conversation about U.S. politics, contacting journalists is a natural strategy to influence news coverage and shape public opinion.   Therefore, as a case study, we collect unseen accounts who have recently contacted high-profile political journalists on Twitter  (Figure \ref{fig:flagged_account_reply}).
%
High-profile journalists were selected from a pre-compiled Twitter list \footnote{\url{https://twitter.com/mattklewis/lists/political-journalists?lang=en}} and the Twitter API was used to collect 47,426 mentions of these 57 journalists, resulting in in 39,103 unique accounts on which we could apply our model.  These accounts represent out-of-sample data for the model.


\section{Method}
Next, we describe how our classifiers were constructed, and then how we performed human evaluation on the classifiers' predictions on unseen, out-of-sample Twitter accounts.
\subsection{Constructing Classifiers}
Our goal is to build classifiers that can detect potential Russian trolls still active on Twitter.
In order to characterize user accounts during classification, we used features that can be grouped into 5 broad categories.

\noindent \textbf{Profile features.} Considering that the Russian trolls are likely to have more recently created Twitter accounts \cite{zannettou2018let}, we calculated the \textit{time since creation} by counting the number of days since a Twitter account's creation date up to January 1, 2019. We also hypothesized there would be a difference in profile descriptions since it requires human effort to customize one's profile \cite{badawy2018analyzing}. Thus, we calculated the \textit{length of profile (characters)}.
Additionally, previous research has shown that Russian trolls tend to follow a lot of users, probably to increase the number of their followers~\cite{zannettou2018disinformation}. So, we also calculated the \textit{number of followers}, \textit{number of following}, and the \textit{ratio of followers to following} for each Twitter account. 

\noindent \textbf{Behavioral features.} The behavioral features we computed were broadly in four categories: i) hashtags, ii) mentions, iii) shared links (URLs), and iv) volume and length of (re)tweets (i.e., tweets and retweets). 
First, considering that the Russian trolls used a high number of certain hashtags before and after the election \cite{zannettou2018let}, we hypothesized there would be a difference in the usage of hashtags and calculated the \textit{average number of hashtags (words)} and \textit{average number of hashtags (characters)}. 
Next, we calculated the \textit{average number of mentions (per tweet)}, as Russian trolls tend to mention more unique users compared to a randomly selected set of normal users \cite{zannettou2018disinformation}.
In order to capture the Russian trolls' behaviors regarding the sharing of links (URLs) in (re)tweets, we also calculated the \textit{average number of links (per tweet)}, \textit{ratio of retweets that contain links among all tweets}, and \textit{ratio of tweets that contain links among all tweets}.
Prior research has shown that temporal behaviors such as retweet rate (i.e., the rate at which accounts retweet content on Twitter) are useful in identifying online campaigns \cite{ghosh2011entropy}. Therefore, we calculated the \textit{average number of tweets (per day)}, \textit{standard deviation of number of tweets (per day)}, and \textit{ratio of number of retweets out of all tweets (retweet rate)} for measuring (re)tweet volume and rate. Additionally, we calculated the \textit{average number of characters of tweets} for obtaining tweet length.

\noindent \textbf{Stop word usage features.} 
Prior research has shown that non-native English speakers use function words, such as articles, and highly-frequent content words (e.g,. ``good'') at different dates than native speakers~\cite{ionin2008sources,nicolai2014does}, which allows recovering their originally spoken language even when the speaker is fluent.   We hypothesized that the linguistic cues in function words enable identifying Russian trolls due to their non-native language skills.   We used the list of 179 stop words provided by the Python library sklearn\footnote{\url{https://github.com/scikit-learn/scikit-learn/blob/master/sklearn/feature_extraction/stop_words.py}} to calculate their frequency rates~\cite{pedregosa2011scikit}; this set is a superset of function words and common content words typically used but allows for maximum reproducibility by others.  For each stop word, we calculated the ratio of the number of each stop word's occurrences among all words present in all tweets by a Twitter account.

\noindent \textbf{Language distribution features.} Additionally, we computed the distribution of languages used by a Twitter account, by examining the account's timeline.
For example, if 5\% of all tweets within a user's timeline were written in \textit{Russian}, then the feature value for Russian would be calculated as 0.05.
A total of 82 languages were identified when computing 
the intersection of all distinct languages used, after combining all the tweets made by Russian troll accounts and control accounts.
In order to identify the language for tweets, we used the Google CLD2  library.

\noindent \textbf{Bag of Words features.} Finally, we tokenized the tweets (converted to lowercase), and broke the text contained in each tweet into words, discarding stopwords (which are captured above). Using these words, we extracted unigrams and bigrams from the text, and constructed a vocabulary for all tweets, containing only the top 5000 most frequent n-grams.
During the classification phase, each tweet can be represented as  a  feature  vector  of  all  words  and  phrases (n-grams) present in the vocabulary (i.e., a Bag of Words (BoW) model), and the feature values would be the frequency of occurrence.


\subsubsection{Classifiers.}
We ran classification tests using Logistic Regression, Decision Tree (maximum depth 10), and Adaptive Boosted Decision Tree, using all 5 categories of features described above.
In order to understand the importance of each individual category of features, we built 5 additional classifiers, one for each category of features, as shown in Table~\ref{table:model_results_each_group}.

\subsection{Testing on Unseen Data \& Human Evaluation} \label{human_eval_section}
While it's interesting to model data previous to 2019 from Twitter, a core question is: \textit{How does this model perform on present-day Twitter?} In order to answer this question, we applied the computational models described above to unseen, out-of-sample Twitter accounts (the journalists' mentions), and manually assessed the models' predictions. 

\subsubsection{Out-of-sample Twitter accounts.}
For further analysis, we picked the Logistic Regression classifier trained on \textit{all features}, as Logistic Regression's simplicity seems most likely to generalize to out-of-sample data.
We ran this classifier on 39,103 unique (unseen) accounts that mentioned 57 high-profile political journalists on Twitter (see Figure \ref{fig:flagged_account_reply} for an example), in order to flag suspicious accounts that resemble Russian trolls. We use the term \textbf{"flagged accounts"} for the accounts among the journalists' mentions that were classified as a potential Russian troll by the model we built, and \textbf{"unflagged accounts"} for the accounts that were not classified as such. 
However, we have \textit{left the world of ground truth}, since there is no hard evidence or labels available to prove or disprove the classifiers' prediction regarding an unseen Twitter account resembling a Russian troll.

\subsubsection{Validation through human evaluation.}
As a result, we included an additional validation step, and employed human evaluation in order to assess the \textit{quality} of classifier predictions, regarding unseen Twitter accounts that mentioned high-profile journalists. 
Out of all the potential Russian troll accounts that were \textit{flagged} by the classifier, we randomly selected 50 of these flagged accounts for the \textit{manual validation} step. In addition to these 50 flagged accounts, we also included 50 \textit{unflagged} accounts, which were also randomly selected for human evaluation. 
These 100 Twitter accounts were combined in random order and manually annotated by three trained raters, one of which also had Political Science expertise.  Their task was to independently rate how likely it was for a given Twitter account to be a Russian troll.
Annotators were not told how many accounts were flagged by the classifier in order to reduce confirmation bias \cite{nickerson1998confirmation}.

\textit{Training for the manual inspection.} Before the human evaluation was conducted, the three raters spent over 12 hours, during a course of 4 weeks, examining the tweets made by Russian trolls, that was officially released by Twitter (dataset described in Section~\ref{russian_troll_tweets}). The raters examined the tweet's content as well as hashtags, links, mentions in the tweets and the trolls' language usage.

\textit{Execution of the manual inspection.}
The raters were instructed to go to each account's profile page and independently study the profile/background image, description, account creation date, etc., and examine the most recent 20 tweets from each of these 100 accounts, to decide how likely it was for each account to be a Russian troll.
The raters focused on characteristics like: 1) frequent (re)tweeting and replying, causing the account to have an abnormal number of tweets compared to the accounts' age, 2) default or vague profiles, 3) whether the account was created recently, 4) behavior of (re)tweeting only political content, 5) no evidence of personal life in its profile page, and 6) extreme statements and inflammatory hashtags.
Working independently, each rater marked their decision as a rating from \textit{1 to 5}, where \textit{5} meant \textit{``highly likely the account is a Russian troll''}, along with writing a brief description of the reasoning behind their decision (e.g., what made the account (un)likely to be a Russian troll?). 

After all three raters completed the evaluation, we calculated the agreement level among their ratings using Krippendorff’s $\alpha$ \cite{krippendorff2011computing}. Finally, we computed the average rating for each account, to decide whether an account flagged by the classifier was indeed verified to be a potential Russian troll, through manual inspection.
The average ratings, and the predictions from the classifier were now shown to the three raters, and they gathered to discuss their observations regarding the characteristics of flagged accounts.


\section{Results}
Next, we describe the results of in-domain (2016 election troll dataset) classification, a SAGE analysis of discriminative words~\cite{eisenstein2011sparse}, and finally the results of human evaluation on out-of-sample data.

\subsection{Cross-validation}
To evaluate our models we first performed 10-fold cross validation. In measuring performance, we prioritized \textit{precision} since we want to minimize false positives considering their potential impact (i.e., flagging a normal account as a suspicious account has a large downside). As shown in Table \ref{table:all_features_model_results}, when using all features, including those from Bag of Words, we see the simplest model, the Logistic Regression classifier, achieving 78.5\% in precision and 98.9\% in terms of AUC. Adaptive Boosted Decision Trees achieves over 85\% for all metrics with the precision being 94.2\% and AUC being 91.2\%. Although one might think the baseline here is about 98.6\% considering just predicting "not troll" for all accounts, accuracy is not the right metric since the dataset is extremely unbalanced.  

We also built classifiers using each category of features in order to see which features are more crucial. As shown in Table \ref{table:model_results_each_group}, the classifier built using \textit{5,000 Bag of Word features} performs overall the best, achieving 98\% in terms of AUC and 58\% in terms of precision. We can also see each classifier built using \textit{language distribution features} and \textit{stop word usage features} both reach above 80\% in terms of AUC.

\begin{table}[t!]
\resizebox{0.48\textwidth}{!}{
\begin{tabular}{l ccccc}
\textbf{Algorithm} & \textbf{Precision} & \textbf{Recall} & \textbf{F1} & \textbf{AUC} & \textbf{Accuracy} \\ 
\midrule
LR & 0.785 &  0.835 & 0.809 & 0.989& 0.994 \\ 
DT & 0.911 &  0.775 & 0.837 &  0.926 & 0.996 \\ 
ADT &  0.942 & 0.884 & 0.912 & 0.912  & 0.998 \\
\end{tabular}
}
\caption{Average performance of the three classifiers using all features to predict from full data using 10-fold cross validation.}
\label{table:all_features_model_results}
\end{table}

\begin{table}[t!]
\centering
\resizebox{0.48\textwidth}{!}{
\begin{tabular}{ r c c c c c } 
 &  \textbf{Profile} & \textbf{Behavior} & \textbf{Language} & \textbf{Stop word}  & \textbf{BoW} \\
 \cmidrule{2-6}
Precision & 0.07 & 0.24 & 0.15 & 0.14 & 0.58 \\ 
 Recall  & 0.01 & 0.4 & 0.21 & 0.39 &  0.79\\ 
F1 &  0.02 & 0.3 & 0.17 & 0.21 & 0.66 \\ 
AUC & 0.74 & 0.76 & 0.85 & 0.86 &  0.98\\ 
Accuracy  & 0.98 & 0.97  & 0.97 & 0.96 &  0.99\\ 
\end{tabular}
}
\caption{Average performance of the Logistic Regression classifiers using each category of features when conducting 10-fold cross validation.}
\label{table:model_results_each_group}
\end{table}

\subsection{SAGE Analysis}
As shown in Table \ref{table:model_results_each_group}, the BoW features were important predictors.
Next, we automatically identified keywords in the vocabulary that distinguished Russian troll accounts from the rest. In order to do so, we used the most recent 200 tweets per
Russian troll accounts (the same tweets used to construct our models). As a baseline comparison, we also compiled the set of tweets made by the control accounts (not identified to be Russian troll accounts). Our goal was to identify terms whose frequencies are especially large in the tweets made by Russian troll accounts, in comparison to tweets by the control group.

\begin{figure}[!t]
    \centering
    \includegraphics[width=0.7\columnwidth]{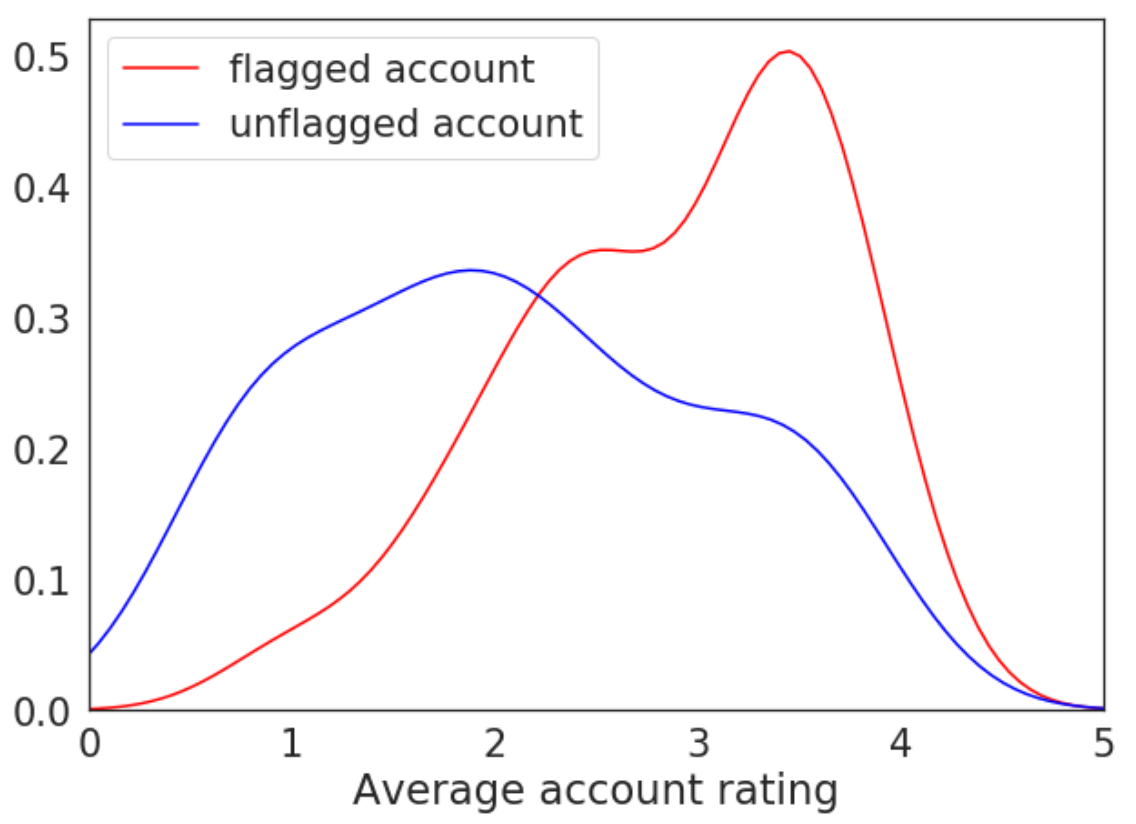}
    \caption{Distributions of average ratings given by 3 raters per account (flagged versus unflagged).}
    \label{fig:average_ratings}
\end{figure}
\begin{table}[!t]
\small
\centering
\begin{tabular}{ l | r } 
\midrule
 \# of Journalists &  57  \\ 
\# of Journalists having at least 1 flagged account &  57 (100\%)\\ 
\midrule
\# of Unique accounts among the mentions& 39,103 \\ 
\# of Unique flagged accounts & 1,466 (3.7\%) \\ 
\midrule
\end{tabular}
\caption{Description of the results when testing the model on unseen accounts that mentioned high-profile journalists on Twitter.}
\label{table:model_flag_results}
\end{table}

\begin{table*}[!t]
\tiny
\centering
\begin{tabular}{ l r r r r r r|r r r r r r r } 

& \textbf{n-gram} & \textbf{SAGE} & \begin{tabular}[c]{@{}c@{}}\textbf{Treatment} \\ \textbf{Count}\end{tabular} & \begin{tabular}[c]{@{}c@{}}\textbf{Treatment} \\ \textbf{Rate}\end{tabular} & \begin{tabular}[c]{@{}c@{}}\textbf{Base} \\ \textbf{Count}\end{tabular} & \begin{tabular}[c]{@{}c@{}}\textbf{Base} \\ \textbf{Rate}\end{tabular} & & \textbf{n-gram} & \textbf{SAGE} & \begin{tabular}[c]{@{}c@{}}\textbf{Treatment} \\ \textbf{Count}\end{tabular} & \begin{tabular}[c]{@{}c@{}}\textbf{Treatment} \\ \textbf{Rate}\end{tabular} & \begin{tabular}[c]{@{}c@{}}\textbf{Base} \\ \textbf{Count}\end{tabular} & \begin{tabular}[c]{@{}c@{}}\textbf{Base} \\ \textbf{Rate}\end{tabular}  \\
\midrule
1 & fukushima2015 & 12.19 & 12498& 0.0068& 1 & 5.14e-09 &
16 & rising & 3.22& 1017& 0.0005& 7004 &3.60e-05 \\
2& fukushimaagain & 11.87 & 8215 & 0.0045 & 0 & 0&
17& quotes & 3.19 & 1075& 0.0006& 7635& 3.92e-05  \\
3 & ukraine & 7.36 & 9327 & 0.0051& 1018 & 5.23e-06 &
18& government& 3.17& 2543& 0.0014& 18579& 9.55e-05\\
4 & \foreignlanguage{russian}{на}& 6.79 & 13622 & 0.0074 & 2657 &1.36e-05 &
19& politics& 3.14 & 1488 & 0.0008 & 11177&5.74e-05   \\
5 & \foreignlanguage{russian}{не}& 6.46 & 8799& 0.0048& 2373 & 1.22e-05 &
20&dangerous& 3.07 & 1390& 0.0007& 11154& 5.73e-05\\
6 & nuclear & 5.70 & 6781& 0.0037& 3922& 2.02e-05&
21& awful& 3.02& 1462& 0.0008& 12380& 6.36e-05 \\
7 & danger & 4.90 & 2822 & 0.0015 & 3644 & 1.87e-05 &
22& quote& 2.98& 2334& 0.0013& 20581& 1.05e-04 \\
8& disaster & 4.74 & 3427& 0.0019 & 5187 &  2.67e-05&
23& emergency & 2.94 & 911& 0.0005& 8351 & 4.29e-05\\
9 & plant & 4.14 & 2697& 0.0015& 7400& 3.80e-05 &
24& hillary& 2.89 & 1716& 0.0009 &16541 & 8.50e-05 \\
10 & louisiana & 4.04 & 2031 & 0.0011& 6170& 3.17e-05&
25&hillary clinton &2.82 &544 &0.0003 & 5585&2.87e-05 \\
11 & turkey & 3.64& 2314& 0.0013& 10509& 5.40e-05&
26&happy thanksgiving &2.78 & 671& 0.0004& 7174&  3.69e-05 \\
12 & officials & 3.54 & 1320& 0.0007& 6642&3.41e-05 &
27& clinton& 2.71& 1328& 0.0007& 15260&7.84e-05  \\
13& crisis & 3.52& 1539& 0.0008& 7906&  4.06e-05&
28& attacks& 2.69& 527& 0.0003 & 6176 & 3.17e-05  \\
14 & midnight & 3.40 & 2859& 0.0015& 16513& 8.49e-05&
29& obama &  2.63 & 3194 & 0.0017 & 39733 & 2.04e-04 \\
15 & rap & 3.32 & 2575 & 0.0014& 16193 & 8.32e-05&
30& maga& 2.61 & 485 & 0.0003 & 6181& 3.18e-05\\
\midrule
\end{tabular}
\caption{Top 30 n-grams distinguishing tweets by Russian troll accounts (treatment group) from control accounts (base group) obtained through SAGE analysis. \textit{SAGE} denotes the score obtained for an n-gram through SAGE analysis---n-grams with higher SAGE scores are more distinguishing of the Russian troll accounts when compared to the control accounts. \textit{Treatment} and \textit{base count} denote the \textbf{raw frequency count} of an n-gram within the treatment and base groups respectively, while \textit{treatment} and \textit{base rate} denote the \textbf{frequency rate} of an n-gram within the treatment and base groups respectively.}
\label{table:SAGE_results}
\end{table*}

Due to the long-tail nature of word frequencies~\cite{zipf1949human}, straightforward comparisons often give unsatisfactory results. The difference in word frequencies between two groups is usually dominated by \textit{stopwords}: a 1\% difference in the frequency of ``the'' or ``it'' will be larger than the overall frequency of most terms in the vocabulary. The ratio of word frequencies---equivalent to the difference in log frequencies, and to pointwise mutual information---has the converse problem: without carefully tuned smoothing, the resulting keywords will include only the lowest frequency terms, suffering from high variance. SAGE offers a middle ground, selecting keywords by comparing the parameters of two logistically-parametrized multinomial models, using a self-tuned regularization parameter to control the tradeoff between frequent and rare terms~\cite{eisenstein2011sparse}. SAGE has been used successfully for the analysis of many types of language differences, including age and gender~\cite{pavalanathan2015confounds}, and politics~\cite{sim2013measuring}.

Thus, we conducted SAGE analysis on the text contained in the tweets made by the two groups of accounts, namely Russian trolls and control accounts.
The top 30 words that distinguished the English-speaking Russian trolls from the control accounts are shown in Table~\ref{table:SAGE_results}. 
We found that the use of keywords associated with chaos and distrust such as \textit{danger}, \textit{disaster}, \textit{crisis}, \textit{emergency}, and \textit{attacks} distinguished Russian troll accounts. We also found that Russian troll accounts used the names of political figures such as \textit{hillary clinton} and \textit{obama} more frequently compared to the non-troll accounts. Additionally, Russian troll accounts used words such as \textit{maga} (Make America Great Again), a term often used by conservatives, and keywords related to the Fukushima disaster, such as \textit{fukushima2015}.

\subsection{Testing on Unseen Data \& Human evaluation}

As shown in Table \ref{table:model_flag_results}, the Logistic Regression classifier trained on all features flagged 1,466 (3.7\%) of the 39,103 accounts mentioning journalists as potential Russian trolls, an average of 25.7 per journalist. All 57 journalists had at least one account flagged as a potential Russian troll by the classifier.

We then asked our three raters to independently indicate how likely they thought 100 sample accounts from the set of account who mentioned journalists were to actually be Russian trolls. These 100 sample accounts were a mix of 50 randomly selected flagged accounts and 50 randomly selected unflagged accounts. Whether each account was flagged or not was not communicated to the raters. The Krippendorff’s $\alpha$ for the three sets of ratings was computed to be 0.512, indicating that the \textit{agreement level among labelers} was moderate \cite{krippendorff2011computing}. This is expected considering this is a very difficult annotation task since the Russian trolls are intentionally trying to blend in. However, this moderate agreement level at distinguishing suspected troll accounts from those of normal users indicates there is some salient behavioral difference between the two groups. Based on the human evaluations, the median rating for all accounts (out of 5.0) provided by the three raters was 2.5 with the median for the flagged accounts being 3.12. 50\% (25/50) of the flagged accounts scored above 3 on average, indicating that they closely resembled Russian troll accounts (Figure \ref{fig:average_ratings}). 70\% (i.e., 35/50) of the flagged accounts had at least 2 raters giving them a score above 3, with 84\% (42/50) of them had at least 1 rater giving a score above 3. This is strong evidence for classification generalization, as it is similar to the in-sample precision scores (these figures correspond to out-of-sample precision).


\section{Analysis of the Flagged Accounts}
Next, we examine the bot scores of the model-flagged accounts, as well as other quantitative covariates, such as deletion and suspension rates.

\subsection{Botometer Scores}
An interesting question we can ask is whether the flagged accounts are just \textit{bot accounts} because previous disinformation campaigns have been coordinated through bots \cite{ferrara2017disinformation}. Researchers at Indiana University provide a tool to the public called Botometer\footnote{\url{https://botometer.iuni.iu.edu/#!/}} that will assess an account for bot characteristics ~\cite{davis2016botornot}. There are user-based features, friends features, network features, temporal features, content and language features, and sentiment features used to classify each account \cite{varol2017online}. 

\subsubsection{Method.} For the potential trolls, we use the 1,466  accounts flagged by the Logistic Regression classifier constructed by using all features. We randomly selected 1,500 users from the unflagged accounts among the mentions for the comparison group. For each group we used the Python Botometer API\footnote{\url{https://github.com/IUNetSci/botometer-python}} to get each account's overall bot scores as well as scores by only using the features per each category.

\subsubsection{Results.} 
\begin{figure}[t!]
    \centering
    \includegraphics[width=0.8\columnwidth]{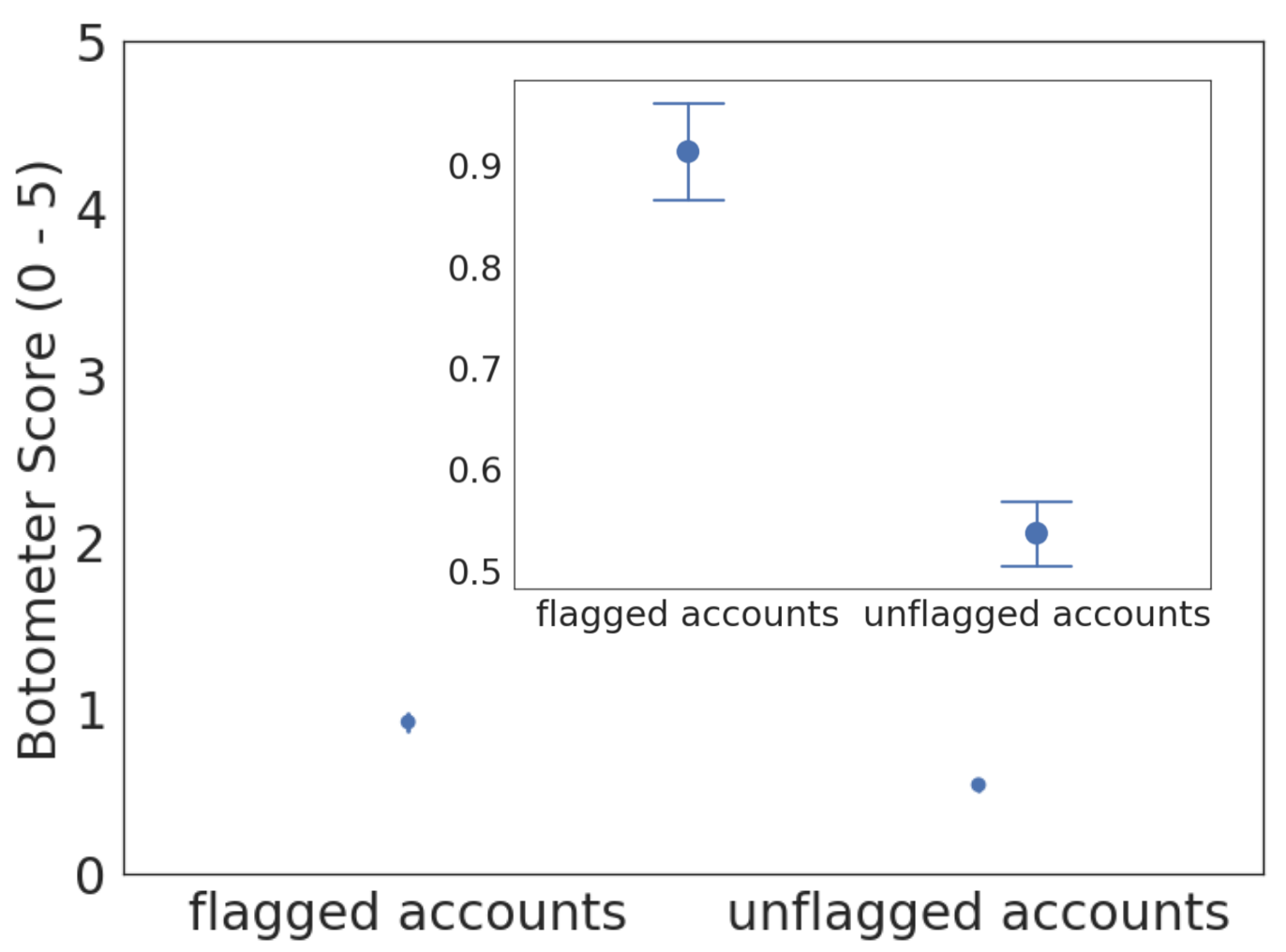}
    \caption{Flagged and unflagged accounts have Botometer scores under 1 (though flagged accounts score higher at a statistically significant level). Inset shows a zoomed in y-axis of larger plot. Bars show 95\% confidence intervals.}
    \label{fig:bot_score_distribution}
\end{figure}

While the flagged accounts do exhibit some bot-like characteristics, it seems that it is not a dominant behavior. 
As shown in Figure \ref{fig:bot_score_distribution}, although the average bot score of the accounts flagged as potential Russian trolls is higher than the unflagged accounts', both of them are less than 1 (on a 5-point scale).

This lack of difference between the two groups would seem essential to both defeat bot-detection software already running on Twitter, as well as provide authenticity in conversations. One working hypothesis is that the Russian trolls might be \emph{software-assisted human workers,} making it hard to detect them with just bot-detection algorithms.

When we take the 6 categories into account, we see that all categories' scores for the flagged accounts are all higher than the unflagged ones, but still lower than 2 (Table \ref{table:bot_score_category_results}). Comparing each category's score of flagged accounts and unflagged accounts shows not much difference, with the highest difference being when only using user-based features (0.49).

\begin{table}[t!]
\small
\centering
\begin{tabular}{ l l l } 
 & \textbf{Flagged} & \textbf{Unflagged} \\
\midrule
User-based &1.3 ($\sigma$=1.25)& 0.81 ($\sigma$=0.97)\\ 
Friends & 1.35 ($\sigma$=0.95)& 1.02($\sigma$=0.67)\\
Network & 1.33  ($\sigma$=1.01)& 1.06 ($\sigma$=0.79)\\
Temporal & 1.42 ($\sigma$=1.25)&  1.08 ($\sigma$=0.99)\\
Content \& Language & 1.34 ($\sigma$=1.22)& 0.94 ($\sigma$=0.91)\\
Sentiment & 1.22 ($\sigma$=1.21)& 0.88 ($\sigma$=0.92)\\
\midrule
\midrule
All & 0.91 ($\sigma$=0.95)& 0.54 ($\sigma$=0.61)\\
\midrule
\end{tabular}
\caption{Comparison of flagged and unflagged accounts' average bot scores calculated using each category and all categories. Highest score is 5.}
\label{table:bot_score_category_results}
\end{table}

\subsection{Flagged Accounts' Behavioral Characteristics}
Next, we highlight interesting quantitative covariates among the flagged accounts. We believe these findings show that the flagged accounts do reasonably resemble Russian trolls.

\noindent \textbf{Deletion or suspension of accounts.} As of January 13, 2019, 4 flagged accounts were suspended while 8 flagged accounts showed up as not existing (the model was run on the accounts before January 13 and these accounts existed at that time).

\noindent \textbf{Account information.} In Table \ref{table:comparison_table} we can see that on average the flagged accounts had more followers and following that unflagged accounts. Russian trolls had more number of followers and following than the control accounts as well (Table \ref{table:dataset}). The account age was calculated by getting the number of days since account creation up to January 1st, 2019. We can see that the flagged accounts are relatively newly created compared to the unflagged ones, which is also a characteristic of Russian trolls (cf. Table \ref{table:dataset}).

\noindent \textbf{Tweet volume and rate.}
The flagged accounts (41.99) tweeted more on a daily basis compared to unflagged ones (26.83) with a higher standard deviation, as shown in Table \ref{table:comparison_table}.

\begin{table}[t]
\small
\begin{tabular}{@{}l|l|l@{}}
\textbf{}              & \textbf{Flagged} & \textbf{Unflagged} \\\midrule
Total \# of Account & 1,441 &1,485 \\ 
Avg Account Age (days)& 1202.43& 2175.25\\ 
Avg \# of Followers & 2205.49 & 1621.7\\ 
Avg \# of Following &1772.6 & 1361.57\\\midrule
Total \# of Tweets & 3,527,171 & 3,515,765 \\
Avg \# of Tweets & 2447.72 & 2268.12\\ 
Avg of avg daily \# of tweets & 41.99 ($\sigma$=24.5)& 26.83 ($\sigma$=16.9)\\
\end{tabular}
\caption{Comparison of flagged and a subset of unflagged accounts regarding account and tweet volume/rate.  }
\label{table:comparison_table}
\end{table}

\noindent \textbf{Language.}
Note that since we ran the model on accounts that recently mentioned high-profile political journalists in U.S., it is likely that both flagged and unflagged accounts use English most frequently. However, we can see that the flagged accounts had more tweets in Russian compared to the unflagged ones (Table \ref{table:top_lang_comparison}). 

\begin{table}[H]
\small
\centering
\begin{tabular}{ l r l r } 
\textbf{Flagged} &  & \textbf{Unflagged} &\\\midrule 
Total tweets & 3,527,171 & Total Tweets  & 3,515,765\\ \midrule
English tweets& 3,368,336 & English tweets& 3,339,469\\
German tweets&  6,736 & Spanish tweets&5,767 \\
Russian tweets& 3,145 & German tweets& 3,833\\
\midrule
\end{tabular}
\caption{Comparison of the top 3 languages used in tweets by the flagged and a subset of unflagged accounts.}
\label{table:top_lang_comparison}
\end{table}

\noindent \textbf{Hashtags.}
As shown in Table \ref{table:top_hashag_comparison}, many of the top hashtags used by the flagged and unflagged accounts overlap with the majority being political ones, such as \#MAGA and \#Trump. However, we can see the flagged accounts use more hashtags overall, such as using \#MAGA more than twice as often as unflagged acccounts, despite the fact that the flagged accounts had slight more tweets.
\begin{table}[t!]
\small
\centering
\begin{tabular}{ l r l r } 
\textbf{Flagged} &  & \textbf{Unflagged} &\\\midrule 
\#MAGA & 12,385 & \#Trump  & 4,553\\
\#TrumpResign & 10,874 &\#MAGA  & 4,499\\
\#Trump &  7,879 & \#TrumpResign & 3,809\\
\#TrumpShutdown & 6,230 & \#TrumpShutdown & 3,047\\
\#Resist & 3,554 & \#ResignTrump & 2,451\\
\#allaboutthemusic & 3,208& \#BREAKING & 2,078\\
\#GOP & 3,089 & \#TrumpShutDown &  1,873\\
\#Kavanaugh &2,358 & \#TheDailyBark  & 1,838 \\
\#TrumpRussia & 2,257 &  \#DogsOfBauwow & 1,793\\
\#labrador  & 2,180 & \#Kavanaugh & 1,660 \\
\midrule
\end{tabular}
\caption{Comparison of top 10 hashtags used by the flagged and unflagged accounts.}
\label{table:top_hashag_comparison}
\end{table}

\section{Discussion}
We have presented results showing that: 1) Russian trolls can be separated from a large control group of accounts active at the same time; 2) that same computational model can uncover likely Russian trolls still active on Twitter today; and 3) that those flagged accounts don't appear to be bots, at least in the way that we currently understand bot behavior. Next, we reflect on observations our raters made while examining flagged accounts, suggesting high-level strategies used by Russian trolls today. We conclude this section with thoughts on implications for social platforms.

\noindent \textbf{Reflections on Flagged Accounts} \label{reflections_on_flagged}
Our three raters closely examined many flagged accounts still active on Twitter today. While their observations are at their core anecdotal, we believe they will be helpful to researchers to identify Russian troll accounts. 

\noindent \textbf{Mismatch of bio and tweets.} Many flagged accounts had discordant bios tweet content. For instance, some flagged accounts had specific universities mentioned in their bio, but none of the tweets the authors examined contained any reference to them. 

\noindent \textbf{Profile picture strategies.}
Profile pictures play important roles in signalling identity on Twitter, and we witnessed various attempts to manage identity through profile pictures among this group as well. For instance, some flagged accounts used a middle-aged white person’s profile pic or a cartoon photo while other accounts had no profile or (and) no cover photo at all. One rater used reverse image engineering to discover that one flagged account claimed to live in Brooklyn, New York, with a profile picture of a restaurant called ``Brooklyn Bar and Grill.'' However, that restaurant is in Iceland. We believe it may be a common strategy to reuse images found elsewhere on the internet as profile photos, as creating unique and believable profile photos is costly.

\begin{figure}[t!]
    \centering
    \includegraphics[width=1\columnwidth]{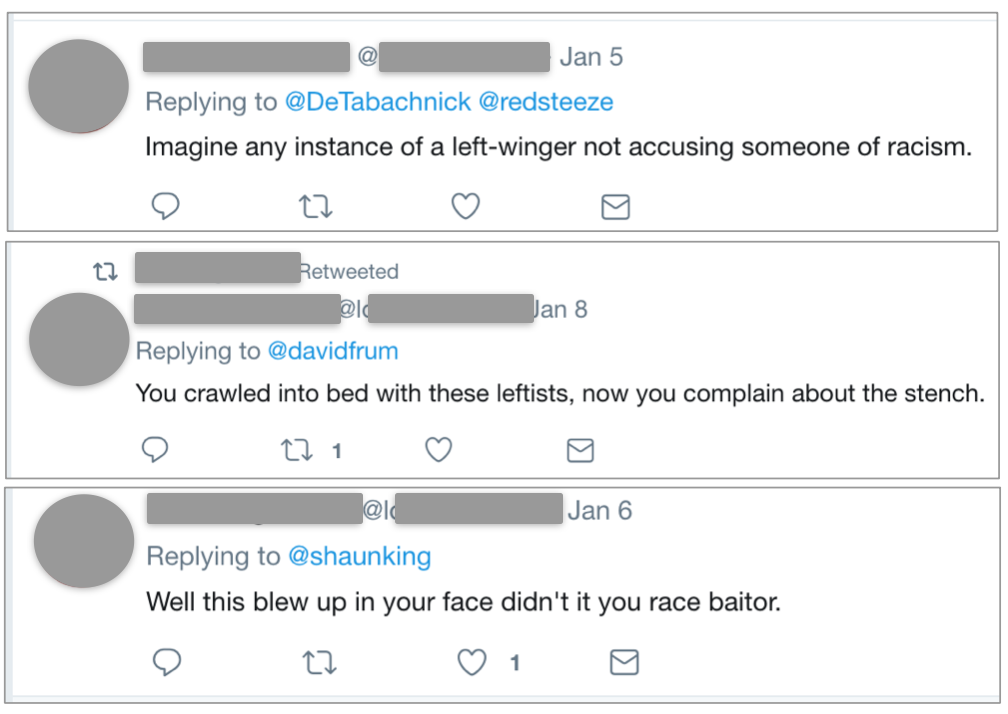}
    \caption{Example of a flagged account replying back to various accounts in an aggressive manner, such as name-calling.}
    \label{fig:ex2}
\end{figure}

\noindent \textbf{Using controversial issues and aggressive behavior.} Based on the tweets we examined, some of the flagged accounts showed aggressive behavior. This included calling names and tweeting about controversial issues like immigration or gun control. An example is depicted in Figure \ref{fig:ex2}.

\noindent \textbf{Abnormal (re)tweet rate.} The raters noted that many of the flagged accounts showed high (re)tweet rate of mostly political content. For instance, one account has been tweeting roughly 65 times per day. For example, one of the flagged accounts examined by our raters has tweeted over 44,200 times since January 2017.
\subsection{Implications for Social Platforms}
One key takeaway from our study is that identifying active Russian trolls on Twitter is possible given even limited access to tweets and user data. It follows that social platforms that have access to complete data about users, their content, and their interactions, can identify Russian trolls among their users. We argue that social platforms should take the initiative to build models and tools to police these accounts. As indicated by our Botometer results, current bot-detection efforts are unlikely to sweep up most Russian troll accounts.



\section{Future Work}

\subsection{Social Platforms Other than Twitter}
Research has shown that the Russian trolls' activities were different on different platforms. For instance, campaigns on Facebook targeted interests of African-Americans and Mexican-Americans, while the Twitter campaign focused more on  topics such as general news and politics \cite{spangher2018analysis}.
Thus, it would be interesting to see how our model can be adpated to other social platforms such as Facebook and Reddit and whether our qualitative findings hold on other platforms.

\subsection{Publicly Available Models}
We have made our models available to broader research community in order to spark more research on this and related topics. For instance, using these models we might be able to understand the social networks of trolls---something that cannot be done with the 2016 election dataset's troll accounts, as Twitter deleted them.
Furthermore, we hope that our publicly available models may help other researchers to flag and interrogate accounts with questions such as "What are their objectives?" and "Have they affected users' political beliefs?"
\section{Conclusion}

Due to the societal importance and impact of social media, coordinated efforts by adversarial individuals and organizations, e.g., trolls of the IRA, have the potential to substantially negatively impact society.  Here, we show that Twitter accounts belonging to Russian trolls can be used to identify new suspected troll accounts.  We develop a new machine learning model that in cross-validation tests achieves an AUC of 98.9\%, allowing us to identify and study new trolls to answer unanswered questions about their activity. When testing our model on accounts that recently mentioned prominent journalists---a likely action to influence the conversation about U.S. politics---we find that Russian trolls are likely still active, with 3.7\% of the accounts contacting these journalists being  flagged as suspected Russian trolls.  Our human evaluation suggests that roughly 70\% of these accounts are valid.  
Finally, we argue that the platforms, which have access to internal signals, can succeed at similar or better rates at finding and deactivating similar adversarial actors. 

\balance{}
\bibliography{references}
\bibliographystyle{aaai}

\end{document}